\documentclass{emulateapj}

\def\ms{m~s$^{-1}$}
\def\ks{km~s$^{-1}$}
\def\msini{$M_P\sin{i}$}

\def\vsini{$v\sin{i_\star}$}

\def\mjup{$M_{\rm Jup}$}

\def\lam{$-33.1^\circ$}
\def\lerr{$7.4^\circ$}
\def\vrot{$2.80$}
\def\vroterr{$0.57$}

\begin{document}

\title{A Third Exoplanetary System with Misaligned Orbital and Stellar
  Spin Axes$^1$}

\author{
  John Asher Johnson\altaffilmark{2,3}, 
  Joshua N.~Winn\altaffilmark{4},
  Simon Albrecht\altaffilmark{4}, \\
  Andrew W.~Howard\altaffilmark{5,6},
  Geoffrey W.~Marcy\altaffilmark{5},
  J. Zachary Gazak\altaffilmark{2}
}

\email{johnjohn@ifa.hawaii.edu}

\altaffiltext{1}{Based on data collected at Subaru Telescope, which is
  operated by the National Astronomical Observatory of Japan; the Keck
  Observatory, which is operated as a scientific partnership among
  the California Institute of Technology, the University of
  California, and the National Aeronautics and Space Administration; and
  the UH\,2.2-meter telescope.}
\altaffiltext{2}{Institute for Astronomy, University
  of Hawaii, Honolulu, HI 96822; NSF Astronomy and Astrophysics
  Postdoctoral Fellow} 
\altaffiltext{3}{NSF Astronomy and Astrophysics Postdoctoral Fellow} 
\altaffiltext{4}{Department of Physics, and Kavli Institute for
  Astrophysics and Space Research, Massachusetts Institute  
  of Technology, Cambridge, MA 02139}
\altaffiltext{5}{Department of Astronomy, University of California,
  Mail Code 3411, Berkeley, CA 94720}
\altaffiltext{6}{Townes Postdoctoral Fellow, Space Sciences
  Laboratory, University of California, Berkeley, CA 94720-7450 USA} 

\begin{abstract}
  We present evidence that the WASP-14 exoplanetary system has
  misaligned orbital and stellar-rotational axes, with an angle
  $\lambda = $\lam$ \pm$ \lerr\ between their sky projections. The
  evidence is based on spectroscopic observations of the
  Rossiter-McLaughlin effect as well as new photometric observations.
  WASP-14 is now the third system known to have a significant
  spin-orbit misalignment, and all three systems have
  ``super-Jupiter'' planets ($M_P > 3 $~\mjup) and eccentric orbits.
  This finding suggests that the migration and subsequent orbital
  evolution of massive, eccentric exoplanets is somehow different from
  that of less massive close-in Jupiters, the majority of which have
  well-aligned orbits.
\end{abstract}


\section{Introduction}

Close-in giant planets are thought to have formed at distances of
several AU and then migrated inward to their current locations
\citep{lin96}. The mechanism responsible for the inward migration of
exoplanets is still debated. Some clues about the migration process
may come from constraints on the stellar obliquity: the angle between
the stellar spin axis and the orbital axis. The sky projection of this
angle, $\lambda$, can be measured by observing and interpreting the
anomalous Doppler shift during the transit of a planet, known as the
Rossiter-McLaughlin effect \citep{mclaughlin24, rossiter24, queloz00,
  winn05, ohta05, gaudi07}. Some of the proposed migration pathways
would produce large misalignments (at least occasionally) while others
would preserve the presumably close alignment that characterizes the
initial condition of planet formation.

For example, theories that invoke migration of the planet through
interactions with the gaseous protoplanetary disk predict small
spin-orbit angles, and that initial spin-orbit misalignments and
eccentricities should be damped out
\citep{lin96,moorhead08,lubow01}. On the other hand, impulsive
processes such as close encounters between planets
\citep{chat08,ford08} or dynamical relaxation \citep{juric08} should
drive systems out of alignment.  The Kozai mechanism also produces
large orbital tilts (Wu \& Murray 2003, Fabrycky \& Tremaine 2007).
Ultimately the hope is that the predictions of migration theories can
be compared with an ensemble of measurements of $\lambda$
\citep{fabrycky09}.

In this paper we add the transiting exoplanet WASP-14\,b to the
growing collection of systems for which the projected spin-orbit angle
has been measured. WASP-14 is a relatively bright ($V = 9.75$) F5V
star which was discovered by the Wide-Angle Search for Planets
(SuperWASP) to undergo periodic transits by a Jovian planet every 2.2
days \citep[][hereafter J09]{joshi09}. The planet is among the most
massive of the known transiting exoplanets, with $M_P = 7.3$~\mjup,
and it has a measurably eccentric orbit ($e = 0.090 \pm 0.003$), which
is unusual among the hot Jupiters. J09 also reported a measurement of
the spin-orbit angle, $\lambda = -14^{+21}_{-14}$ degrees, which is consistent
with zero, but also allows for the possibility of a significant
misalignment. In the following section we describe our spectroscopic
and photometric observations of WASP-14, made in an attempt to refine
the measurement of $\lambda$. In \S~\ref{analysis} we present evidence
for a large spin-orbit misalignment based on our radial-velocity
measurements obtained during transit. We summarize the results of our
joint analysis of our photometric and spectroscopic monitoring in
\S~\ref{discussion}, and present tentative evidence of an emerging
trend between spin-orbit misalignment, and the physical and orbital
characteristics of close-in exoplanets.

\section{Observations and Data Reduction}
\label{observations}

\subsection{Radial Velocity Measurements}
\label{sec:rv}

\begin{figure}[!h]
\epsscale{1.2}
\plotone{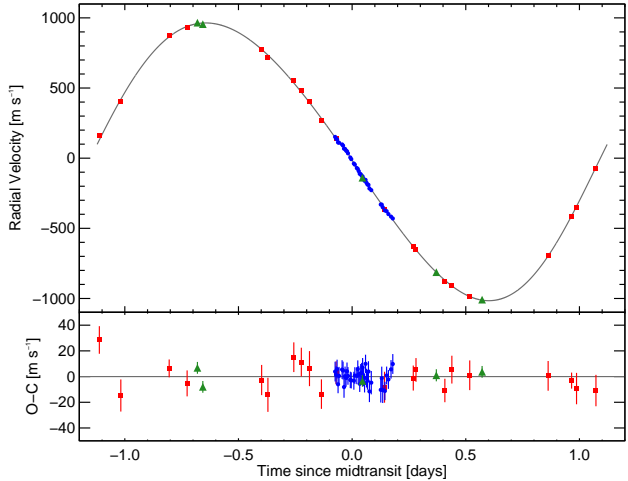}
\caption{ Relative radial velocity measurements of WASP-14 as a
  function of orbital phase, expressed in days since midtransit.  The
  symbols are as follows: Subaru (circles), Keck (squares), Joshi et
  al.\ 2009 (triangles). The lower panel shows the residuals after
  subtracting the best-fitting model including both the Keplerian
  radial velocity and the Rossiter-McLaughlin
  effect. \label{fig:orbit}}
\end{figure}

We observed the transit predicted by J09 to occur on 2009~June~17
using the High-Dispersion Spectrometer \citep[HDS,][]{noguchi02} on
the Subaru 8.2\,m Telescope atop Mauna Kea in Hawaii. We obtained
spectra of WASP-14 through an iodine cell using the I2b spectrometer
setting and a 0\farcs8 slit, providing a resolution of approximately
$60,000$. We started our observing sequence just after evening
twilight, about 20~min before the predicted time of ingress. We
continued our observations until 2.5~hr after egress when the star set
below $20^{\circ}$ elevation. For most of our observations we used
exposure times of 5~min, yielding a signal-to-noise ratio (SNR) of
100--120~pixel$^{-1}$ at 5500~\AA, the central wavelength of the range
with plentiful iodine absorption lines. At high airmass we increased
our exposure times to 10~min.

We also obtained 8 radial velocity measurements of the G2V star
HD\,127334 on the same night, 2009~June~17.  HD\,127334 is a long-term
target of the California Planet Search. Keck/HIRES radial velocity
measurements over the past 3 years show that the star is stable with
an rms scatter of 2.5~\ms.  With HDS we used exposure times ranging
from 60 to 120 seconds, resulting in SNR ranging from
110--120~pixel$^{-1}$ at 5500~\AA.

We also obtained out-of-transit (OOT) radial velocities of WASP-14
using the High-Resolution (HIRES) spectrometer on the Keck I telescope
starting in July 2008. We set up the HIRES spectrometer in the same
manner that has been used consistently for the California Planet
Search \citep{howard09}. Specifically, we employed the red
cross--disperser and used the I$_2$ absorption cell to calibrate the
instrumental response and the wavelength scale \citep{marcy92b}. The
slit width was set by the $0\farcs86$ B5 decker, and the typical
exposure times ranged from 3-10~min, giving a resolution of about
60,000 and a SNR of 140-250~pixel$^{-1}$ at 5500~\AA.

For the spectra obtained at both telescopes, we performed the Doppler
analysis with the algorithm of \citet{butler96}, as updated over the
years. A clear, Pyrex cell containing iodine gas is placed in front of
the spectrometer entrance slit. The dense forest of molecular lines
imprinted on each stellar spectrum provides a measure of the
wavelength scale at the time of the observation, as well as the shape
of the instrumental response \citep{marcy92b}. The Doppler shifts were
measured with respect to a ``template'' spectrum based on a
higher-resolution Keck/HIRES observation from which the spectrometer
instrumental response was removed, as far as possible, through
deconvolution. We estimated the measurement error in the Doppler shift
derived from a given spectrum based on the weighted standard deviation
of the mean among the solutions for individual 2~\AA\ spectral
segments. The typical measurement error was 1.0-1.7~\ms\ for the Keck
data and 6~\ms\ for the Subaru data. The RV data are given in
Table~\ref{tab:vels} and plotted in Figure~\ref{fig:orbit}.

\subsection{Photometric Measurements}

We observed the photometric transit of 2009~May~12 with the University
of Hawaii 2.2~m (UH2.2m) telescope on Mauna Kea. We used the
Orthogonal Parallel Transfer Imaging Camera (OPTIC), which is equipped
with two Lincoln Labs CCID128 orthogonal transfer array (OTA)
detectors \citep{tonry97}. Each OTA detector has 2048$\times$4096
pixels and a scale of 0\farcs135 pixel$^{-1}$.  OTA devices can shift
accumulated charge in two dimensions during an exposure. We took
advantage of this charge-shifting capability to create large
square-shaped point spread functions (PSFs) that permit longer
exposures before reaching saturation \citep{howell03, tonry05,
  johnson09}.

\begin{figure}[!t]
\epsscale{1.2}
\plotone{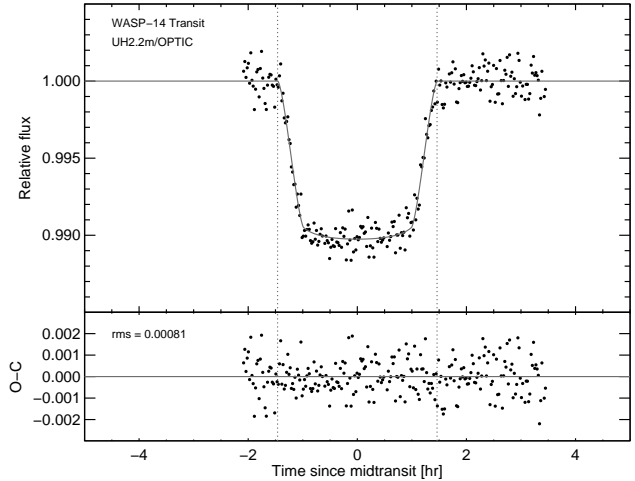}
\caption{{\it Top Panel:} Relative photometry of WASP-14 during the
  transit of 2009~May~12. {\it Bottom Panel:} Residuals from the
  best-fitting transit light curve model. \label{fig:transit}}
\end{figure}

We observed the transit of WASP-14 continuously for 5.5~hr spanning
the transit. We observed through a custom bandpass filter centered at
850~nm with a 40~nm full-width at half-maximum. We shifted the
accumulated charge every 50~ms to trace out a square-shaped region 25
pixels on a side. Exposure times were 50~s, and were separated by a
gap of 29~s to allow for readout and refreshing of the detectors. Bias
subtraction and flat-field calibrations were applied using custom IDL
procedures described by \citet{johnson09}.

The only suitable comparison star that fell within the OPTIC field of
view is a $V=12.1$ star $\sim6$ arcminutes to the Northeast.  The
fluxes from the target and the single comparison star were measured by
summing the counts within a square aperture of 64 pixels on a side.
Most of the light, including the scattered-light halo, was encompassed
by the aperture. We estimated the background from the outlier-rejected
mean of the counts from four rectangular regions flanking each of the
stars \citep{johnson09}. As a first order correction for variations in
sky transparency, we divided the the flux of WASP-14 by the flux of
the comparison star. The transit light curve is shown in
Figure~\ref{fig:transit}, and the photometric measurements and times of
observations (HJD) are listed in Table~\ref{tab:phot}.

\begin{figure*}[!t]
\epsscale{1}
\plotone{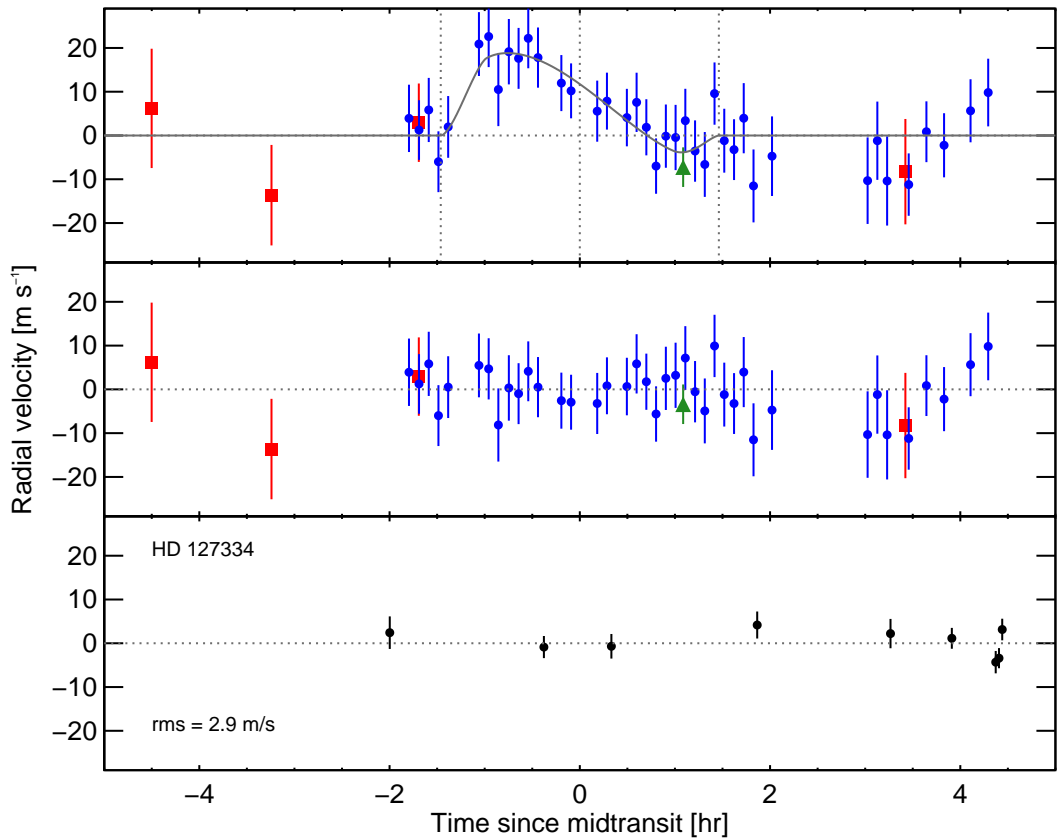}
\caption{Relative radial velocity measurements made during transits of
  WASP-14. The symbols are as follows: Subaru (circles), Keck
  (squares), Joshi et al.\ 2009 (triangles). {\it Top Panel:} The
  Keplerian radial velocity has been subtracted, to isolate the
  Rossiter-McLaughlin effect. The predicted times of ingress,
  midtransit and egress are indicated by vertical dotted lines. {\it
    Middle Panel:} The residuals after subtracting the best-fitting
  model including both the Keplerian radial velocity and the RM
  effect. {\it Bottom Panel:} Subaru/HDS measurements of the standard star
  HD\,127334 made on the same night as the WASP-14 transit.
  \label{fig:lambda}}
\end{figure*}

\section{Data Analysis}
\label{analysis}

\subsection{Updated Ephemeris}
\label{ephem}

The first step in our analysis was to refine the estimate of the
orbital period using the midtransit time derived from our OPTIC light
curve. We fitted a transit model to the light curve based on the
analytic formulas of \citet{mandelagol} for a quadratic limb-darkening
law. The adjustable parameters were the midtransit time $T_t$, the
scaled stellar radius $R_\star/a$ (where $a$ is the semimajor axis),
the planet-star radius ratio $R_P/R_\star$, the orbital inclination
$i$, the limb darkening coefficients $u_1$ and $u_2$\footnote{We
  allowed both coefficients to be free parameters, subject to the
  conditions $0 < u_1+u_2 < 1$ and $u_1 > 0$.}, and two parameters $k$
and $m_0$ describing the correction for differential airmass
extinction. The airmass correction was given by
\begin{equation}
m_{\rm cor} = m_{\rm obs} + m_{0} + kz
\end{equation}
\noindent where $m_{\rm obs}$ is the observed instrumental magnitude,
$z$ is the airmass, and $m_{\rm cor}$ is the corrected magnitude that
is compared to the transit model \citep{winn09b}.

We used the rms of the OOT measurements as an estimate for the
individual measurement uncertainties. We did not find evidence for
significant time-correlated noise using the time-averaging method of
\citet{pont06}. We fitted the light curve model and estimated our
parameter uncertainties using a Markov Chain Monte Carlo
algorithm \citep[MCMC;][]{tegmark04, ford05, gregory05}. The results for
the midtransit time, and the refined orbital period (using the new
midtransit time and the midtransit time given by J09) are
\begin{eqnarray}
T_c & = & 2454889.8921 \pm  0.00025 \\
P   & = & 2.2437704 \pm    0.0000028~{\rm \,days}.
\end{eqnarray}
The other derived lightcurve parameters were consistent with those
reported by J09.

\subsection{Evidence for Spin-Orbit Misalignment: A Simple Analysis}
\label{simple}

Figure~\ref{fig:transit} shows our Subaru/HDS and Keck/HIRES RV
measurements made near transit, after subtracting the best-fitting
Keplerian orbital model. The RVs measured just after ingress are
redshifted with respect to the Keplerian orbital velocity.  We
interpret this ``anomalous'' redshift as being due to the blockage by
the planet of the blueshifted limb of the rotating stellar surface. We
therefore conclude that the planet's orbit is prograde.

In addition, the anomalous redshift persists until about 1~hr
after midtransit. This is evidence for a misalignment between the
orbital axis and stellar rotation axis.  Were $\lambda=0$, the
midpoint of the transit chord would be on the projection of the
stellar rotation axis, and therefore the anomalous Doppler shift would
vanish at midtransit, in contradiction of the data.  Thus we can
conclude that the orbit of WASP-14\,b is inclined with respect to the
projected stellar spin axis. In the next section we make a
quantitative assessment of $\lambda$.

\subsection{Global Analysis of Radial Velocities and Photometry}
\label{global}

We simultaneously fitted a parametric model to our OPTIC light curve
and to the the 4 sets of RV data: Subaru/HDS and Keck/HIRES (this
work), and OHP/SOPHIE and NOT/FIES (J09). To make our analysis of the
RM effect largely independent of J09, we did not include the RVs
gathered by J09 during transits. The photometric aspects of the model
were given in \S~\ref{ephem}.  The RV model was the sum of the radial
component of the Keplerian orbital velocity, and the anomalous
velocity due to the Rossiter-McLaughlin effect. To compute the latter,
we used the ``RM calibration'' procedure of Winn et al.~(2005): we
simulated spectra exhibiting the RM effect at various orbital
phases\footnote{For the template spectrum, which should be similar to
  that of WASP-14 but with slower rotation, we used a Keck/HIRES
  spectrum of HD\,3681 \citep[$T_{eff} = 6220$~K, {\rm
    [Fe/H]}$=+0.08$; \vsini~$ = 2.8 \pm 0.5$~\ks][]{valenti05}.}, and
then measured the anomalous 
radial velocity $\Delta V_R$ of the simulated spectra using the same
algorithm used on the actual data. We found the results to be
consistent with the formula
\begin{equation}
\Delta V_R = -(\delta f) v_p
\left[1.124 - 0.395 \left(\frac{v_p}{3.5~{\mathrm{km~s}}^{-1}}\right)^2 \right],
\end{equation}
where $\delta f$ is the instantaneous fractional loss of light during
the transit and $v_p$ is the radial velocity of the occulted portion
of the stellar disk.

\begin{figure}[!t]
\epsscale{1.2}
\plotone{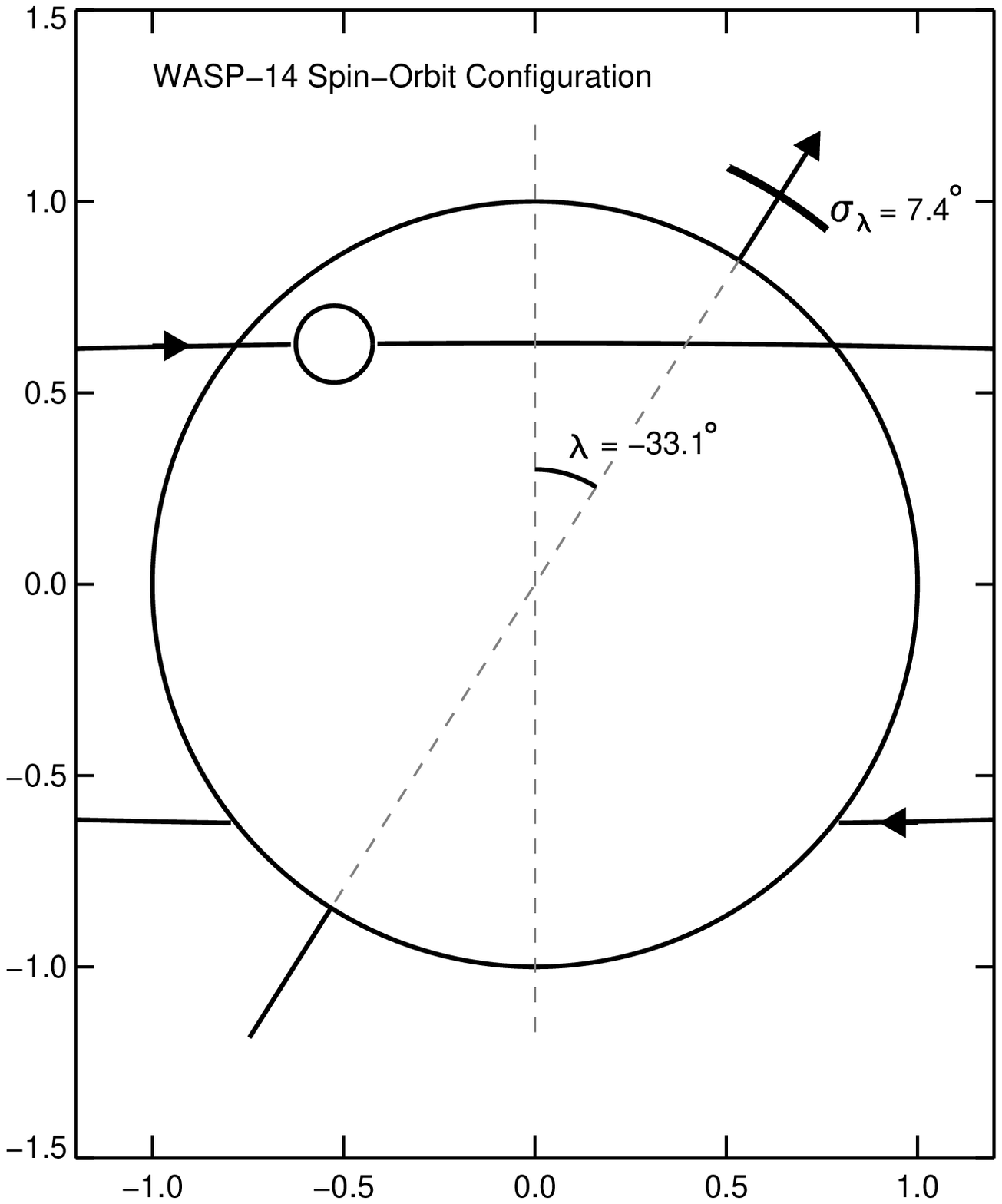}
\caption{The spin-orbit configuration of the WASP-14 planetary
  system. The star has a unit radius and the relative size of the
  planet and impact parameter are taken from the best-fitting transit
  model. The sky-projected angle between the stellar spin axis
  (diagonal dashed line) and the planet's orbit normal (vertical
  dashed line) is denoted by $\lambda$, which in this diagram is measured
  counter-clockwise from the orbit normal. Our best-fitting $\lambda$
  is negative. The 68.3\% confidence interval for $\lambda$ is traced
  on either side of the stellar spin axis and denoted by
  $\sigma_\lambda$.  \label{fig:star}}
\end{figure}

The 18 model parameters can be divided into 3 groups. First are the
parameters of the spectroscopic orbit: the period $P$, the midtransit
time $T_t$, the radial-velocity semiamplitude $K$, the eccentricity
$e$, the argument of pericenter $\omega$, and velocity offsets for
each of the 4 different groups of RV data. Second are the photometric
parameters: the planet-to-star radius ratio $R_p/R_\star$, the orbital
inclination $i$, the scaled stellar radius $R_\star/a$ (where $a$ is
the semimajor axis), the 2 limb-darkening coefficients, and the
out-of-transit flux and differential extinction coefficient. Third are
the parameters relevant to the RM effect: the projected stellar
rotation rate $v\sin i_\star$ and the angle $\lambda$ between the sky
projections of the orbital axis and the stellar rotation axis [for
illustrations of the geometry, see Ohta et al.~(2005), Gaudi \& Winn
(2007), or Fabrycky \& Winn (2009)].

The fitting statistic was
\begin{eqnarray*}
\chi^2 & = &
\sum_{j=1}^{247}
\left[
\frac{f_j({\mathrm{obs}}) - f_j({\mathrm{calc}})}{\sigma_{f,j}}
\right]^2
\\ & + &
\sum_{j=1}^{64}
\left[
\frac{v_j({\mathrm{obs}}) - v_j({\mathrm{calc}})}{\sigma_{v,j}}
\right]^2
\\ & + &
\left[ \frac{P - 2.243770~{\rm d}}{0.0000028~{\rm d}} \right]^2
,
\end{eqnarray*}
where $f_j$(obs) are the relative flux data from the OPTIC light curve
and $\sigma_{f,j}$ is the out-of-transit rms.  Likewise $v_j$(obs) and
$\sigma_{v,j}$ are the RV measurements and uncertainties.  For
$\sigma_{v,j}$ we used the quadrature sum of the measurement error and
a ``jitter'' term of 4.4~m~s$^{-1}$, which was taken from the
empirical calibration of Wright~(2004). The final term enforces the
constraint on the orbital period based on the new ephemeris described
in the previous section.

As before, we solved for the model parameters and uncertainties using
a Markov Chain Monte Carlo algorithm. We used a chain length of
$5\times 10^6$ steps and adjusted the perturbation size to yield an
acceptance rate of $\sim$40\%. The posterior probability distributions
for each parameter were approximately Gaussian, so we adopt the median
as the ``best--fit'' value and the standard deviation as the
1-$\sigma$ error. For the joint model fit the minimum $\chi^2$ is
291.4 with 295 degrees of freedom, giving $\chi^2_\nu = 0.99$. The
contributions to the minimum $\chi^2$ from the flux data and the RV
data were 246.1 and 45.3, respectively. The relatively low value of
the RV contribution compared to the number of RV data points (64)
suggests that 4.4~m~s$^{-1}$ is an overestimate of the jitter for this
star, and that consequently our parameter errors may be slightly
overestimated, but to be conservative we give the results assuming a
jitter of 4.4~m~s$^{-1}$.

\begin{figure*}
\epsscale{1}
\plotone{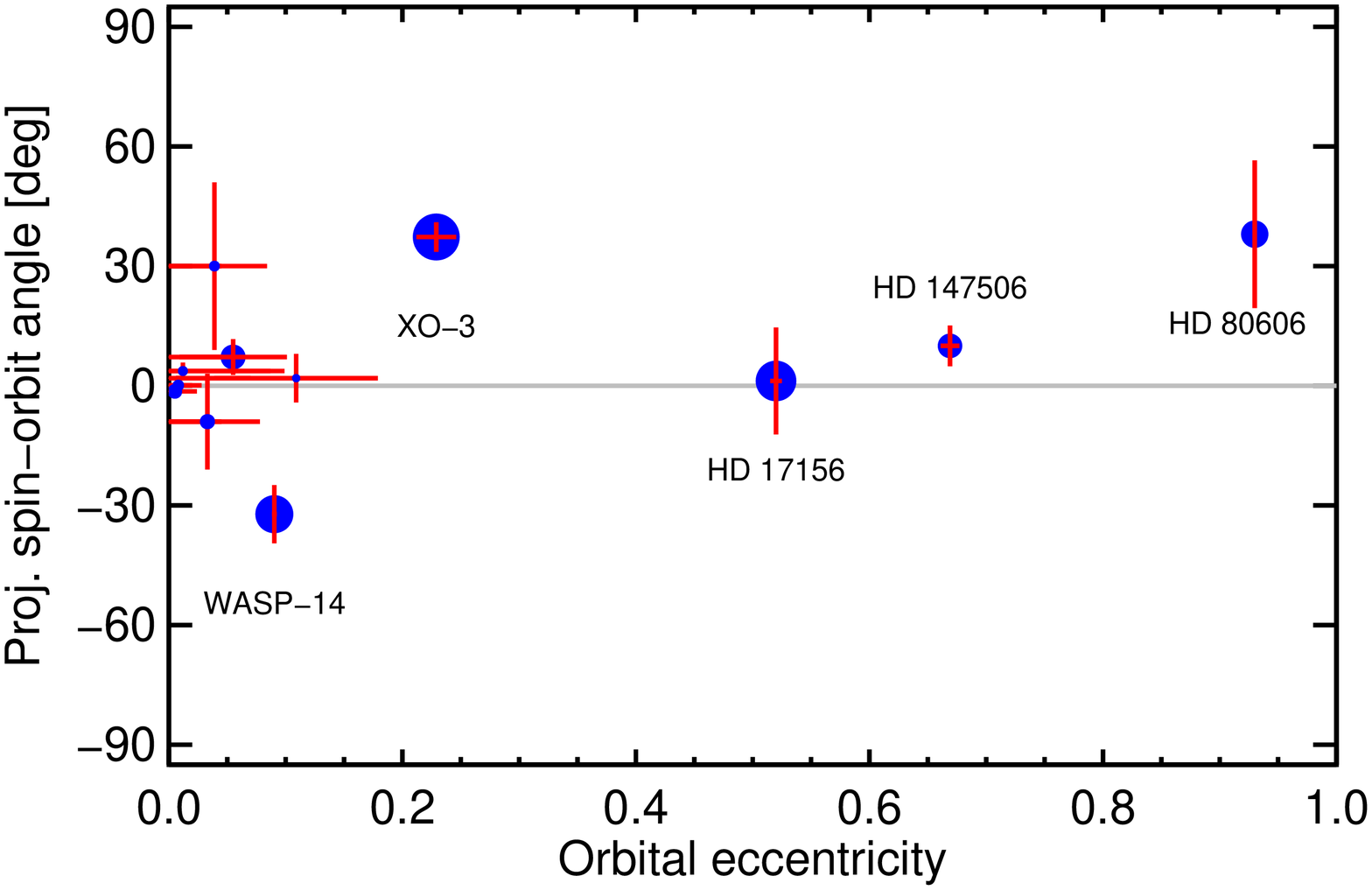}
\caption{The eccentricities and projected spin-orbit angles for the 13
  transiting systems for which the Rossiter-McLaughlin effect has been
  observed and modeled. The data are from Table~1 of
  \citet{fabrycky09} and \citet{mad08} as updated by
  \citet{winn09b}, \citet{narita09} and this work. The size of the
  plot symbols scales as $M_P^{1/2}$, and the error bars show the
  measurement uncertainties for both eccentricity and the spin-orbit
  angle. For the systems consistent with $e=0$ the horizontal error
  bar gives the 95.4\% confidence upper limit on $e$, and for the
  other systems the horizontal error bars represent the 68.3\%
  confidence uncertainties. \label{fig:rmdist}}
\end{figure*}

For the main parameter of interest, the projected spin-orbit angle,
our analysis gives $\lambda = $~\l~$\pm$~\lerr
(Figure~\ref{fig:lambda}). Thus, the WASP-14 planetary system is
prograde and misaligned, as anticipated in the qualitative discussion
of \S~\ref{simple}. Our measurement of $\lambda$ agrees with the value
measured by J09 ($-14^{+21}_{-14}$ deg), but with improved precision
that allows us to exclude $\lambda = 0$ with high confidence. We find
the projected stellar rotational velocity to be $v \sin{i_\star} =
$~\vrot~$ \pm $~\vroterr~\ks. This value is somewhat lower than, but
consistent with, the values determined by J09 from line broadening $v
\sin{i_\star} = 3.0 \pm 1.5$~\ks, from their RM analysis $v
\sin{i_\star} = 4.7 \pm 1.5$~\ks, and from our SME analysis $v
\sin{i_\star} = 3.5 \pm 0.5$~\ks. This agreement among the rotation
rates provides a consistency check on our analysis. The best-fitting
parameters and their uncertainties are listed in Table~\ref{tab:pars}

We found no evidence for another planet or star in the WASP-14 system.
To derive quantitative constraints on the properties of any distant
planets, we added a single new parameter $\dot{\gamma}$ to our model,
representing a constant radial acceleration. A third body with mass
$M_3 \ll M_\star$, orbital distance $a_3 \gg a$ and inclination $i_3$
would produce a typical radial accleration
\begin{equation}
\dot{\gamma} \sim \frac{G M_3 \sin i_3}{a_3^2},
\end{equation}
and our result is $\dot{\gamma}=2.0\pm 1.4$~cm~s$^{-1}$~d~$^{-1}$, or
$1.01\pm 0.72$~$M_{\rm Jup}$~(5~AU)$^{-2}$.

\section{Summary and Discussion}
\label{discussion}

We present new photometric and spectroscopic measurements of the
WASP-14 transiting exoplanetary system. By combining a new transit
light curve, several Keck/HIRES RV measurements made outside of
transit, and most importantly, Subaru/HDS RVs spanning a transit, we
have measured and interpreted the RM effect. By modeling the RM
anomaly we find that the projected stellar spin axis and the planetary
orbit normal are misaligned, with $\lambda = $~\lam$\pm$~\lerr.

Of the 13 transiting systems with measured spin-orbit angles, only 3
have clear indications of spin-orbit misalignments.  The other two
cases besides WASP-14 are XO-3 (H\'ebrard et al.~2008, Winn et
al.~2009) and HD~80606 \citep[][; Winn et al. 2009c in
prep]{gillon09b, pont09}. It is striking that all 3 tilted systems
involve planets several times more massive than Jupiter that are on
eccentric orbits, and that none of the systems with eccentricities
consistent with circular or
with masses smaller than 1~$M_{\rm Jup}$ show evidence for
misalignments (Figure~\ref{fig:rmdist}).

In addition to the three known super-Jupiters with inclined orbits,
there are also two eccentric, massive exoplanets with small projected
spin-orbit angles: HD\,17156\,b
\citep{fischer07,barb07,cochran08,narita09} and HD\,147506
\citep{bakos07b, winn07a,lo08}. However, neither of these cases
presents as strong an exception to the pattern as it may seem. The
measurement of $\lambda$ in both cases was hampered by the poor
constraint on the transit impact parameter, which causes a strong
degeneracy between $\lambda$ and $v\sin i$ (Gaudi \& Winn~2007). It
should also be kept in mind that the measured quantity $\lambda$ is
only the {\it sky-projected}\, spin-orbit angle, and that the true
angle of one or both of those systems may have a stellar rotation axis
that is inclined by a larger angle along our line of sight.

It was already known that the orbits of massive planets are
systematically different from the orbits of less massive planets. For
example, \citet{wright09}, building on a previous finding by
\citet{marcy05a}, showed that planets with minimum masses \msini~$>
1$~\mjup\ typically have lower orbital eccentricities than those with
minimum masses smaller than 1~\mjup. While sub-Jovian-mass planets
have eccentricities that peak near $e = 0$ with a sharp decline toward
$e = 0.4$, those with \msini~$ > 1$~\mjup\ have eccentricities that
are uniformly distributed between $e = 0$ and 0.55.

The tendency for misaligned orbits to be found among massive planets
on eccentric orbits does not yet have a clear interpretation. It may
seem natural for inclined orbits and eccentric orbits to go together,
since both inclinations and eccentricities can be excited by few-body
dynamical interactions, whether through the Kozai effect,
\citep{fab07, wu07} planet-planet scattering \citep[see,
e.g.,]{juric08}, or scenarios combining both of these phenomena
(Nagasawa et al.~2008). However, the mass dependence of these and
other mechanisms for altering planetary orbits needs to be clarified
before any comparisons can be made to the data.

The misalignment of the WASP-14 planetary system, along with the
previously discovered misaligned systems, have offered a tantalizing
hint of an emerging trend among the orbital and physical properties of
close-in, transiting exoplanets. However, trends seen in small data
sets can often be misleading. To bring this picture into better
focus, a more sophisticated analysis of the extant data, following the
example of \citet{fabrycky09} may be better. And as with all
astrophysical trends, observations of a larger sample of objects will
provide a much clearer picture than any statistical analysis
of a smaller sample. Thus, additional RM observations of transiting
systems are warranted, with particular attention paid to trends with
orbital eccentricity and planet mass.

\acknowledgements We thank the referee, Dan Fabrycky, for a remarkably
timely and helpful review. We gratefully acknowledge the assistance of
the UH 2.2~m telescope staff, including Edwin Sousa, Greg Osterman and
John Dvorak. Special thanks to John Tonry for his helpful discussions
and comprehensive 
instrument documentation for OPTIC, Debra Fischer for her HDS
raw reduction code, and Scott Tremaine for his helpful comments and
suggestions. JAJ is an NSF Astronomy and 
Astrophysics Postdoctoral Fellow with support from the NSF grant
AST-0702821. JNW thanks the NASA Origins of Solar Systems program for
support through awards NNX09AD36G and NNX09AB33G, as well as the
support of the MIT Class of 1942 Career Development Professorship. SA
acknowledges support by a Rubicon fellowship from the Netherlands
Organisation for Scientific Research (NWO). We also appreciate funding from
NASA grant NNG05GK92G (to GWM), and AWH gratefully acknowledges
support from a Townes Postdoctoral Fellowship at the UC Berkeley Space
Sciences Laboratory. The authors wish to extend special
thanks to those of Hawaiian ancestry on whose sacred mountain of Mauna
Kea we are privileged to be guests. Without their generous
hospitality, the observations presented herein would not have been
possible.

\bibliography{}

\clearpage

\begin{deluxetable}{lrrc}
\tablecaption{Radial Velocity Measurements of WASP-14\label{tab:vels}}
\tablewidth{0pt}
\tablehead{
\colhead{Heliocentric Julian Date (HJD)} &
\colhead{RV} &
\colhead{Uncertainty} & 
\colhead{Telescope\tablenotemark{a}} \\
\colhead{} &
\colhead{(m~s$^{-1}$)} &
\colhead{(m~s$^{-1}$)} &
\colhead{} \\
}
\startdata
2454667.80421 &  -139.4 &  1.0 & K \\
2454672.81824 & -1008.4 &  1.3 & K \\
2454673.83349 &   955.3 &  1.3 & K \\
2454999.76227 &   151.6 &  6.3 & S \\
2454999.76665 &   138.8 &  5.2 & S \\
2454999.77091 &   133.6 &  5.9 & S \\
2454999.77517 &   112.0 &  5.4 & S \\
2454999.77943 &   110.2 &  5.5 & S \\
2454999.79290 &    97.9 &  5.8 & S \\
2454999.79716 &    89.8 &  5.4 & S \\
2454999.80142 &    67.8 &  7.1 & S \\
2454999.80600 &    65.8 &  6.0 & S \\
2454999.81026 &    54.3 &  5.4 & S \\
2454999.81452 &    49.0 &  5.2 & S \\
2454999.81878 &    34.6 &  5.3 & S \\
2454999.82901 &     4.9 &  4.7 & S \\
2454999.83327 &    -6.9 &  4.5 & S \\
2454999.84471 &   -38.3 &  5.4 & S \\
2454999.84898 &   -46.0 &  4.8 & S \\
2454999.85771 &   -70.3 &  4.9 & S \\
2454999.86197 &   -76.8 &  5.1 & S \\
2454999.86623 &   -92.5 &  4.7 & S \\
2454999.87049 &  -111.3 &  4.5 & S \\
2454999.87477 &  -114.5 &  5.7 & S \\
2454999.87904 &  -124.9 &  6.0 & S \\
2454999.88330 &  -131.0 &  5.8 & S \\
2454999.88757 &  -148.0 &  5.5 & S \\
2454999.89183 &  -161.0 &  6.0 & S \\
2454999.89610 &  -154.8 &  5.6 & S \\
2454999.90037 &  -175.6 &  5.8 & S \\
2454999.90464 &  -187.6 &  5.4 & S \\
2454999.90890 &  -190.4 &  6.7 & S \\
2454999.91317 &  -215.8 &  7.1 & S \\
2454999.92129 &  -228.0 &  8.0 & S \\
2454999.96315 &  -330.4 &  8.8 & S \\
2454999.96743 &  -331.1 &  7.8 & S \\
2454999.97170 &  -350.1 &  9.2 & S \\
2454999.98117 &  -372.5 &  5.6 & S \\
2454999.98891 &  -378.0 &  5.4 & S \\
2454999.99665 &  -398.5 &  5.9 & S \\
2455000.00825 &  -416.6 &  5.7 & S \\
2455000.01598 &  -429.6 &  6.4 & S \\
2455014.86287 &   965.8 &  1.4 & K \\
2455015.91393 &  -812.7 &  1.7 & K \\
\enddata
\tablenotetext{a}{
K: HIRES, Keck~I 10m telescope, Mauna Kea, Hawaii. 
S: HDS, Subaru 8m telescope, Mauna Kea, Hawaii. 
}
\end{deluxetable}

\begin{deluxetable}{cl}
\tablecaption{Relative Photometry for WASP-14\label{tab:phot}}
\tablewidth{0pt}
\tablehead{
\colhead{Heliocentric Julian Date (HJD)} &
\colhead{Relative Flux} \\
}
\startdata
2454963.85021 & 1.00064 \\
2454963.85113 & 1.00127 \\
2454963.85204 & 1.00024 \\
2454963.85296 & 1.00086 \\
2454963.85387 & 1.00115 \\
2454963.85478 & 0.99986 \\
2454963.85569 & 1.00183 \\
2454963.85660 & 1.00005 \\
... & ...
\enddata
\tablecomments{The full version of this table is available in the
  online edition, or by request to the authors.}
\end{deluxetable}

\begin{deluxetable}{lc}
\tablecaption{System Parameters of WASP-14\label{tab:pars}}
\tablewidth{0pt}
\tablehead{
  \colhead{Parameter} & 
  \colhead{Value}     \\
}
\startdata
\emph{Orbital Parameters} & \\
Orbital period, $P$~[days]                & $2.2437704 \pm 0.0000028$  \\
Mid-transit time, $T_t$~[HJD]             & $2454963.93676 \pm 0.00025$  \\
Velocity semiamplitude, $K_\star$~[\ms]    & $989.9 \pm 2.1$  \\
Argument of pericenter, $\omega$~[degrees] & $253.10 \pm 0.80$ \\
Orbital eccentricity, $e$                  & $0.0903 \pm 0.0027$ \\
Velocity offset, $\gamma_{\mathrm{FIES}}$~[\ms]     & $-4989.5 \pm 3.4$ \\
Velocity offset, $\gamma_{\mathrm{SOPHIE}}$~[\ms]     & $-4990.1 \pm 3.0$ \\
Velocity offset, $\gamma_{\mathrm{HIRES}}$~[\ms]  & $107.1 \pm 2.1$ \\
Velocity offset, $\gamma_{\mathrm{HDS}}$~[\ms]    & $7.7 \pm 2.5$ \\
& \\
\emph{Spin-orbit Parameters} & \\
Projected spin-orbit angle $\lambda$~[degrees]  & \lam~$\pm$~\lerr \\
Projected stellar rotation rate $v\sin{i_\star}$~[\ks] & \vrot~$\pm$~\vroterr
\enddata

\end{deluxetable}

\end{document}